\documentclass[a4paper,conference]{IEEEtran}
\IEEEoverridecommandlockouts
\usepackage{subcaption}
\usepackage{cite}
\usepackage{amsmath,amssymb,amsfonts}
\usepackage{algorithmic}
\usepackage{graphicx}
\usepackage{textcomp}
\usepackage{color}
\DeclareMathOperator{\diag}{diag}
\begin{document}
	\title{ 3D Beamforming in Reconfigurable Intelligent Surfaces-assisted Wireless Communication Networks }
	
\author{\IEEEauthorblockN{S. Mohammad Razavizadeh}
	\IEEEauthorblockA{\textit{School of Electrical Engineering} \\
		\textit{Iran University of Science and Technology (IUST)}\\
		Tehran, Iran \\
		smrazavi@iust.ac.ir}
	\and
	\IEEEauthorblockN{ Tommy Svensson}
	\IEEEauthorblockA{\textit{Department of Electrical Engineering} \\
		\textit{Chalmers University of Technology}\\
		Gothenburg, Sweden \\
		tommy.svensson@chalmers.se}
}

	\maketitle
\begin{abstract} Reconfigurable Intelligent Surfaces (RIS)  or Intelligent Reflecting  Surfaces (IRS) are metasurfaces that  can be deployed in various places in wireless environments to make these environments controllable and reconfigurable. In this paper, we investigate  the problem of using 3D beamforming  in RIS-empowered wireless networks and propose a new scheme that provides more degrees of freedom in designing and deploying the RIS-based networks. In the proposed scheme,  a base station (BS) equipped with a full dimensional array of antennas optimizes its radiation pattern in the three dimensional space to maximize the received signal to noise ratio at a target user. We also study the effect of angle of incidence of the received signal by the RIS on its reflecting properties and find a relation between this angle and the BS antenna array's tilt and elevation angles. The user receives the signal from a reflected path from the RIS as well as from a direct path from the BS which both depend on the BS antenna array's tilt  and elevation angles. These angles and also the RIS element's phase shifts are jointly numerically optimized. Our simulation results show that using RIS-assisted 3D beamforming with optimized phase shifts and radiation angles   can considerably improve the performance of wireless networks.  \\
\end{abstract}


\section{Introduction}
A common assumption in all  analysis and studies on wireless communication is that the  propagation environment is static and has an uncontrollable behavior.  However, recent studies show that by proper placement of  Reconfigurable Intelligent Surfaces (RISs);  a.k.a., intelligent reflecting surfaces (IRSs), smart surfaces and large intelligent metasurfaces (LIMs); in the environment, the wireless channel can be electronically  controlled to have desired behavior and  enable better performance in wireless systems \cite{Basar2019}. These surfaces are in fact reflecting antennas made from metasurfaces and their reflection properties are controlled by  electronic devices like pin-diodes or varactors.

Recently, RIS-assisted (or IRS-assisted) wireless networks have attracted   much attention from  researchers and different aspects of these networks have been studied in the literature \cite{Basar2019}. In almost all of these papers, it is assumed that an RIS is a passive reflector metasurface composed of a number of reflecting elements (meta-atoms) and by optimizing the phase shifts at these elements, the received signal at the destination can be improved  \cite{Abeywickrama2019, Wu2019, Xu}. These performance improvements can be measured by various metrics like spectral efficiency \cite{Han2019, Nadeem2019AsymptoticAO}, secrecy rate \cite{Chen, Xu} and energy efficiency \cite{Huang2019}.

\begin{figure}[t]
	\centering
	\includegraphics[width=9.1cm]{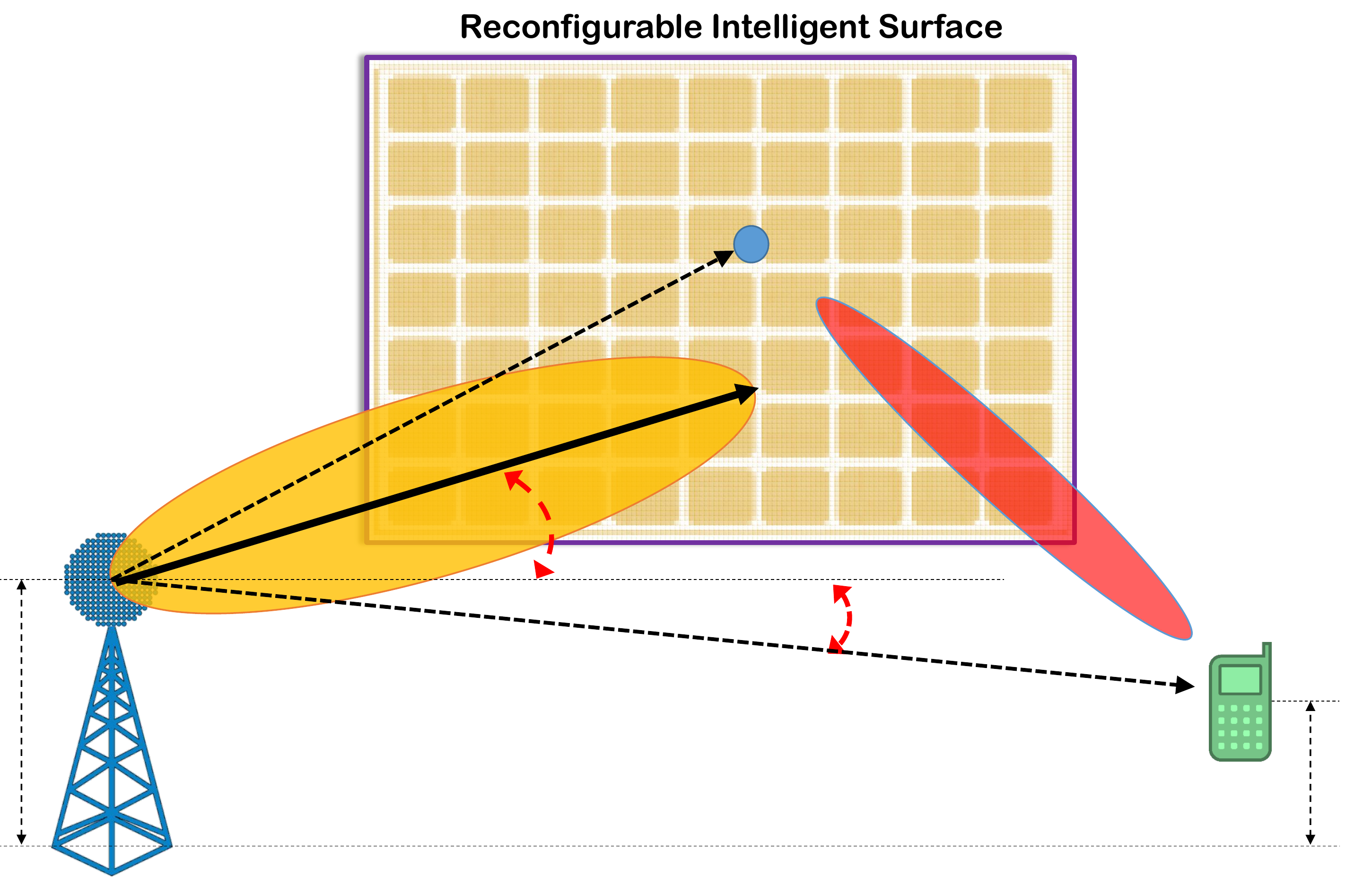}
	\caption{\small{Illustration of an RIS-assisted wireless network showing one RIS-reflected path and  a direct link between the BS and a user.}} 
	\label{FigSysMod1}
\end{figure}

Another technique that has been proposed to be used in the next generation of wireless networks (5G and 6G) is three dimensional (3D) beamforming \cite{Razavizadeh14}. In this technique, the radiating pattern of the base station  (BS) can be controlled in both azimuth and elevation planes in the 3D space. In fact, in addition to the other network parameters, two other design  parameters  are the BS antenna array's tilt and azimuth angles \cite{Khavari}. By adjusting these angles, the radiating beam can be pointed to any desired direction which improves the performance of the network.

In this paper, we  study the  3D beamforming technique in an RIS-assisted wireless network. We consider a network consisting of a multi-antenna BS, an RIS and a single antenna user equipment (UE) in which the UE receives the signal from both a direct path from the BS and also through a reflected path from the RIS. The BS array changes its radiation orientation (i.e.  tilt and azimuth angles) in the 3D space and hence the received signals by the RIS and the UE vary accordingly.  Our main idea is that by adapting the incident angle of the signal received by the RIS, which in turn is controlled by the BS antenna's tilt and azimuth angles, we can have more control of the behavior of the surface and finally more degrees of freedom for the received signal by the UE. The reflected signal by the RIS also depends  on the phase shift of its reflecting elements that need to be optimized. Hence, we formulate an optimization problem that jointly optimizes the radiation angles of the BS as well as the RIS elements' phase shifts to maximize the signal to noise ratio (SNR) at the UE. We numerically solve the resulting optimization problem. The results illustrate the efficiency of the proposed scheme and show that optimizing the radiation angles at the BS in conjunction with optimizing the phase shifts at the RIS can considerably improve the performance of the RIS-assisted wireless networks. In our simulations, we compare the performance of the optimal phase shifts at the RIS with the case that the RIS phases are selected randomly and also the case that the phase shift matrix at the RIS is an identity matrix.   
We also study the effect of the number of reflecting elements at the RIS on the performance of the network. 

The rest of the paper is organized as follows. Section \ref{sectionsystemmodel} describes the system model and 3D modeling of the transmitted signal by the BS. In Sec. III, the received signals at the RIS and the user are derived and the received SNR at the UE is calculated. Section \ref{sectionoptimization} presents the optimization problem for the BS radiation  angles and the RIS phase shifts. Numerical results are presented in Section \ref{sectionsimulation} and finally the paper is concluded in Section \ref{section conclusion}.

\section{System Model}
\label{sectionsystemmodel}
As depicted in Fig. \ref{FigSysMod1}, we consider an RIS-empowered wireless network  in which a BS equipped with an array of  $M$ antennas serves a single antenna user. The channel between the BS and the RIS, between the BS and the UE and between the RIS and the BS are denoted by $\mathbf{H} _r\in \mathbb{C}^{N \times M}$, $\mathbf{h} _d\in \mathbb{C}^{M \times 1}$  and $\mathbf{g}\in \mathbb{C}^{N \times 1}$, respectively. We study a simple system consisting of one RIS and one user in order to gain some basic understanding of RIS-assisted 3D beamforming. 

We assume that the antenna array at the BS is a uniform rectangular array with $M$ antennas placed at the $x$-$z$ plane and  can electronically adjust its radiation beam in a 3D space. The BS antenna's radiation orientation in 3D space is defined by a tilt (vertical or elevation) angle  $\theta_\mathrm{BS} \in [-\pi/2,\pi/2]$ in the vertical domain and an azimuth (horizontal or scan) angle $\phi_\mathrm{BS}\in [-\pi/2,\pi/2]$ in the horizontal domain.  
As depicted in Fig. \ref{Fig3Dmodel},  $\theta_\mathrm{BS}$ is defined as the vertical angle between the BS antenna main beam (boresight angle) and the horizontal axis.  $\phi_\mathrm{BS}$ is also the horizontal angle between the BS antenna boresight and an axis that is normal to the BS array (we assume that this axis is parallel to the RIS).  Utilizing active antenna systems (AAS), these two angles can be controlled electronically without changing the actual mechanical orientation of the BS array. 

The vertical antenna attenuation (pattern) is defined as \cite{Razavizadeh14}
\begin{equation} \label{elev}
A_V \left( \theta_\mathrm{BS} \right)  =  \min {\left[12 \left( \frac{\theta_\mathrm{BS}-\theta_{o}}{\theta_{3dB}} \right)^2 , \ A_m^V \right]},
\end{equation}

\noindent where $\theta_{3dB}$ is the 3dB beamwidth of the BS antenna radiation pattern and $A_m^V $ is the maximum side-lobe level of the antenna pattern in the vertical plane.  The angle $\theta_{o} $ is the direction of the receiver (RIS or UE) in the vertical plane. 

Similarly, in the horizontal  domain, the attenuation in the radiation pattern is defined as \cite{Razavizadeh14}
\begin{equation} \label{azimuth}
A_H \left( \phi_\mathrm{BS} \right)  =  \min {\left[12 \left( \frac{\phi_\mathrm{BS}-\phi_{o}}{\phi_{3dB}} \right)^2 , \ A_m^H \right]},
\end{equation}
\noindent where $\phi_{3dB}$ is the 3dB (half power) beamwidth of the BS antenna radiation pattern and $A_m^H $ is the maximum side-lobe level of the antenna pattern in the horizontal domain.  The angle $\phi_{o} $ is the direction of the receiver (RIS or UE) in the horizontal plane.

Using \eqref{elev} and \eqref{azimuth}, the overall BS antenna gain in dB scale becomes \cite{Nadeem}
\begin{equation} \label{gaindB}
A \left( \theta_\mathrm{BS} , \phi_\mathrm{BS} \right)  = A_{max} - \min\left( A_V \left( \theta_\mathrm{BS} \right)+A_H \left( \phi_\mathrm{BS} \right), A_m \right).
\end{equation}

\noindent Here, $A_{max}$ is the maximum directional gain of the antenna array elements. In addition, it is usually assumed that $A_m=A_m^V = A_m^H =\infty$.  From (1), (2) and (3), and by defining  $\alpha_{max} = 10^{A_{max}/10}$, the BS antenna gain in linear scale becomes 

\begin{equation} \label{gain}
\alpha \left( \theta_\mathrm{BS} , \phi_\mathrm{BS} \right)  =  \alpha_{max} \ 10^{- 1.2 \left[  \left( \dfrac{\theta_\mathrm{BS}-\theta_{o}}{\theta_{3dB}} \right)^2 + \left( \dfrac{\phi_\mathrm{BS}-\phi_{o}}{\phi_{3dB}} \right)^2 \right]   },
\end{equation}
\section{Received Signal at the RIS and UE} \label{sectionUErcv}
\begin{figure}[t]
	\centering
	\begin{subfigure}[b]{0.45\textwidth}
		\centering
		\includegraphics[width=.93\textwidth]{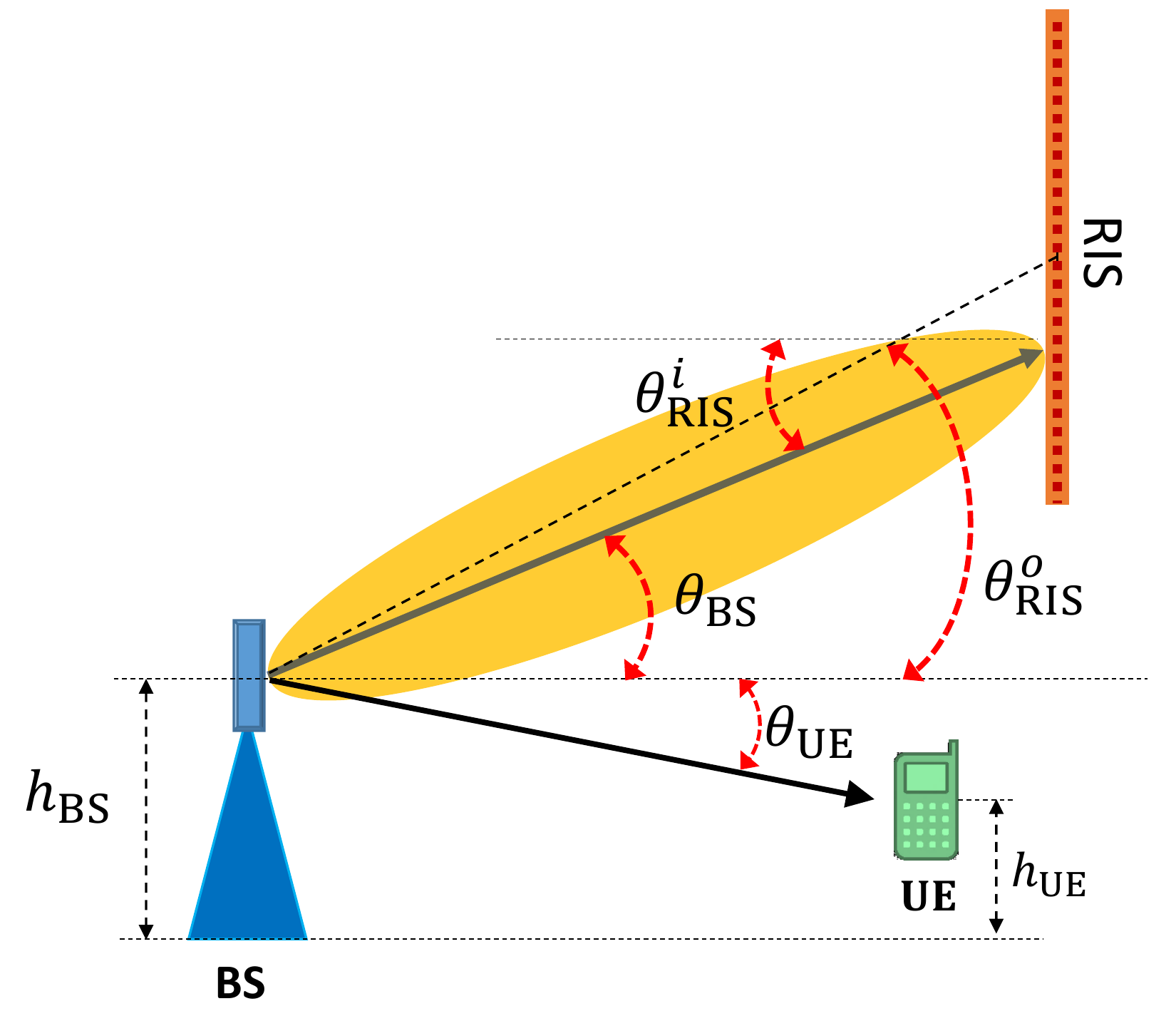}
		\caption{Vertical angle ($\theta$) view}
		\label{fig2vert}
	\end{subfigure} 
	\hfill
	\begin{subfigure}[b]{0.45\textwidth}
		\centering
		\includegraphics[width=.93\textwidth]{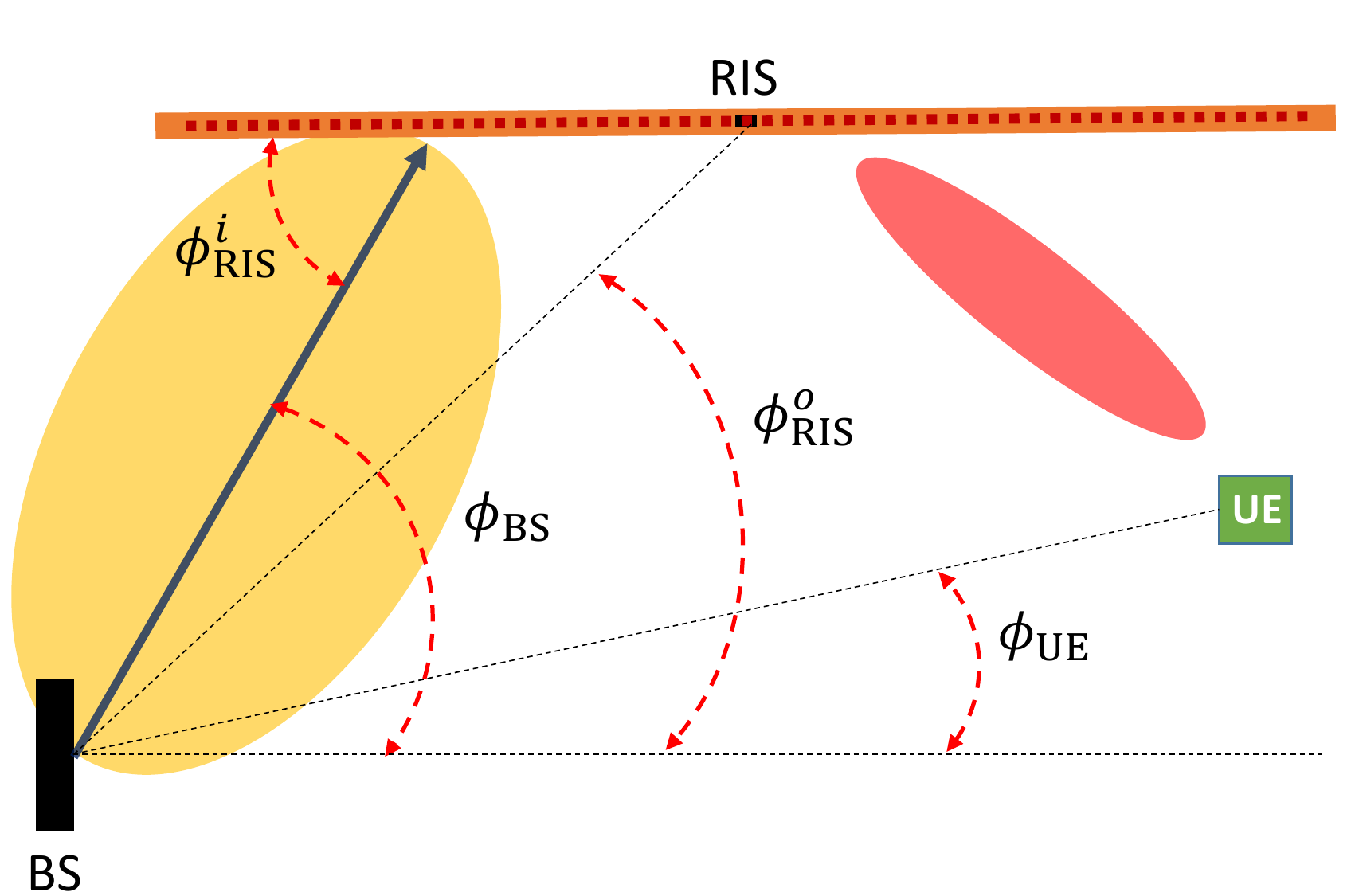}
		\caption{Horizontal angle ($\phi$) view}
		\label{fig2hor}
	\end{subfigure}   
	\caption{\small{3D dimensional modeling of the network.}}
	\label{Fig3Dmodel}
\end{figure}

 \ \\Assuming  the transmitted signal by the BS is denoted by $s$, the received signal at the RIS becomes
\begin{equation} \label{RISReceivedSig}
\mathbf{y}_R  = \sqrt{\alpha^r \left( \theta_\mathrm{BS} , \phi_\mathrm{BS} \right)} \ \mathbf{H}_r \mathbf{w}   s ,
\end{equation}
\noindent where $\mathbf{w} \in \mathbb{C}^{M \times 1}$ is the beamfoming vector at the BS, and 
\begin{equation} \label{gainR}
\begin{split}
& \alpha^r   =  \alpha_{max} \ 10^{- 1.2 \left[  \left( \dfrac{\theta_\mathrm{BS}-\theta_\mathrm{RIS}^o}{\theta_{3dB}} \right)^2 + \left( \dfrac{\phi_\mathrm{BS}-\phi_\mathrm{RIS}^o}{\phi_{3dB}} \right)^2 \right]   }
\end{split}.
\end{equation}
 The angles $\theta_\mathrm{RIS}^o$ and $\phi_\mathrm{RIS}^o$ are defined as in Fig. \ref{Fig3Dmodel}  and we assume that they are all the same among the RIS elements. In other words, although different elements of the RIS receive signals from different angles,  $\theta_\mathrm{RIS}^o$ and $\phi_\mathrm{RIS}^o$ can be assumed as the average of these angles and are used among all elements.   Also, we assume that the transmitted signal is normalized, i.e., $E\lbrace | s |^2 \rbrace  = 1$.

The RIS  consists of $N$ elements where the $i$'th element, $i=\left\lbrace 1,2,...,N\right\rbrace $,  introduces a phase $\psi_i \in [0,2\pi ]$ 
and amplitude $\beta_i \in [0,1 ]$  to the reflected signal.
The overall reflection matrix  of the RIS is defined as 
\begin{equation}
\label{RISmodel1}
\mathbf{\Omega} (\mbox{\boldmath$\beta$}, {\Psi})= \diag\left( \beta_1 e^{j \psi_1} ,  \beta_2 e^{j \psi_2} , \ldots,  \beta_N e^{j \psi_N} \right),
\end{equation}
where ${\Psi} =\left[ \psi_1,  \psi_2,\ldots , \psi_N\right] $ and $\mbox{\boldmath$\beta$} =\left[ \beta_1,  \beta_2,\ldots, \beta_N\right] $. 

An important fact about metasurface-based reflector arrays and RISs is that \textit{their reflection properties depend  on the incident angle  of the incoming wave}. The  angle of the incident signal to the RIS can be defined by two angles of  $\theta_\mathrm{RIS}^i$ and $\phi_\mathrm{RIS}^i$ which are defined in Fig.~\ref{Fig3Dmodel}. In other words, similar to the BS, the received signal by the RIS comes from an elevation (or tilt) angle $\theta_\mathrm{RIS}^i$ and an azimuth angle $\phi_\mathrm{RIS}^i$. 

In a more accurate modeling of the RIS behavior, the phase and amplitudes of the elements depend on the incident angle. Therefore, from \eqref{RISmodel1}, the reflection matrix of the RIS can be written as 
\begin{equation}
\label{RISmodel2}
\begin{split}
&\mathbf{\Omega}\left[   \mbox{\boldmath$\beta$}(\theta_\mathrm{RIS}^i,\phi_\mathrm{RIS}^i), \Psi \right] =  \diag [  \beta_1(\theta_\mathrm{RIS}^i,\phi_\mathrm{RIS}^i) e^{j\psi_1}, \\
& \beta_2(\theta_\mathrm{RIS}^i,\phi_\mathrm{RIS}^i) e^{j \psi_2} , \ldots, \beta_N(\theta_\mathrm{RIS}^i,\phi_\mathrm{RIS}^i) e^{j \psi_N}].
\end{split}
\end{equation}
It should be noted again that in \eqref{RISmodel2}, we assume that all the elements in the RIS  have the same incident angle. This is because the distance between the BS and the RIS is assumed to be very large compared to  the dimension of the surface. 

In addition, it is easily seen from Fig.  \eqref{Fig3Dmodel} that 
\begin{equation}
\label{anglesApprox}
\begin{split}
&\theta_\mathrm{RIS} = \theta_\mathrm{BS}  \overset{\Delta}{=} \theta  \\
&\phi_\mathrm{RIS}  = \phi_\mathrm{BS} \overset{\Delta}{=}  \phi.
\end{split}
\end{equation}
\noindent Please note that there is a little difference between the definition of the tilt and azimuth angles in the BS and the RIS. For example, the tilt angle at the BS is defined as an up-tilt angle whereas in the RIS,  $\theta_\mathrm{RIS}$ is a down-tilt angle. Eq. \eqref{anglesApprox} states that the absolute values of these angles are the same. 

From the electromagnetic theory, we know that the maximum reflection from the surface is when the incoming wave is perpendicular to the surface, i.e. when $ \theta =0$ and $ \phi = \pi/2$ in Fig. \ref{Fig3Dmodel}. On the other hand, if the incoming wave is parallel to the surface (i.e. $ \theta =\pi/2 \ \text{or} -\pi/2$ and $ \phi = 0 \ \text{or}  \ \pi$), the minimum induction will happen and the reflected signal of the surface will be very weak.  An example function that can model this bahaviour is as follows. The amplitude of the reflection matrix, i.e. the vector in  $\mbox{\boldmath$\beta$}$ in \eqref{RISmodel1}, can be approximated by
\begin{equation}
\beta_i(\theta,\phi) = K_1 \cos(\theta)  \sin(\phi) + K_2.
\label{beta_sincos}
\end{equation}
In \eqref{beta_sincos}, $K_1$ and  $K_2$ are  two constants that determine the gain of the RIS and since the RISs are usually passive and do not amplify the signal,  $K_1 + K_2=1$ (i.e. the maximum value of $\beta_i$ is one). In addition, $K_2$ shows the reflection amplitudes in the case that the radiated signal is parallel to the RIS and hence it is  much smaller than $K_1$ (for example in our simulations, we assume $K_1=0.9$ and  $K_2=0.1$).  It should be noted that the reflection amplitude $\beta_i$ is the same on all the reflecting elements (i.e. it is independent of $i$).
Substituting \eqref{beta_sincos} in \eqref{RISmodel2}, we have
\begin{equation}
\label{RISmodel3}
\mathbf{\Omega}( \Psi,\theta,\phi) = \left[ K_1 \cos(\theta)  \sin(\phi) + K_2\right] \diag[ e^{j {\psi_1}} , \ldots,  e^{j {\psi_N}}].
\end{equation}
The UE receives signals from a direct path  from the BS and also from a reflected path off the RIS and it can be  written as 
\begin{equation} \label{UEReceivedSig}
y_u   = y_u^d + y_u^R + n.
\end{equation}
Here, $n \sim \mathcal{CN}(0,\sigma^2)$ is the received additive white Gaussian noise at the user and $y_u^d$ is the received signal from the direct path that can be expressed as
\begin{equation} \label{UEReceivedSigD}
y_u^d   = \sqrt{\alpha^d \left( \theta_\mathrm{UE} , \phi_\mathrm{UE} \right)} \ \mathbf{h}_d^T \mathbf{w}   s ,
\end{equation}
\noindent where $\theta_\mathrm{UE}$  and  $\phi_\mathrm{UE}$ are the tilt and azimuth angles of the UE with respect to the BS radiation pattern that are fixed for a given position of the user. Furthermore, 
\begin{equation} \label{gainD}
\begin{split}
& \alpha^d  =  \alpha_{max} \ 10^{- 1.2 \left[  \left( \dfrac{\theta_\mathrm{BS}-\theta_\mathrm{UE}}{\theta_{3dB}} \right)^2 + \left( \dfrac{\phi_\mathrm{BS}-\phi_\mathrm{UE}}{\phi_{3dB}} \right)^2 \right] }.
\end{split}
\end{equation}
 In addition,  $ y_u^R$  is the UE received signal from the RIS. This signal can be expressed as
\begin{equation} \label{UEReceivedSigR}
\begin{split}
y_u^R  & =  \mathbf{g}^H \mathbf{\Omega} \ \mathbf{y}_R  = \sqrt{\alpha^r \left( \theta_\mathrm{BS} , \phi_\mathrm{BS} \right)} \ \mathbf{g}^H  \mathbf{\Omega} \  \mathbf{H}_r \ \mathbf{w} \  s.
\end{split}
\end{equation}
By substituting  \eqref{UEReceivedSigD} and \eqref{UEReceivedSigR} in \eqref{UEReceivedSig}, we have 
\begin{equation} \label{UEReceivedSig2}
\begin{split}
y_u   &= \sqrt{\alpha^d \left(\theta_\mathrm{BS} , \phi_\mathrm{BS} \right)} \ \mathbf{h}_d^H \mathbf{w}   s  \\
&+ \sqrt{\alpha^r \left( \theta_\mathrm{BS} , \phi_\mathrm{BS} \right)} \ \mathbf{g}^H  \mathbf{\Omega} \  \mathbf{H}_r \ \mathbf{w}  s + n.\\
&= \left[ \sqrt{\alpha^d } \ \mathbf{h}_d^H + \sqrt{\alpha^r } \ \mathbf{g}^H  \mathbf{\Omega} \  \mathbf{H}_r  \right]  \mathbf{w} s + n.  \\
\end{split}
\end{equation}
From \eqref{UEReceivedSig2}, the instantaneous SNR of the received signal can be obtained as 
\begin{equation} 
\label{SNR}
\gamma = \frac{   \left|  \left(\sqrt{\alpha^d } \ \mathbf{h}_d^H    
+ \sqrt{\alpha^r} \ \mathbf{g}^H  \mathbf{\Omega} \  \mathbf{H}_r \ \right) \mathbf{w} \right|^2  }{\sigma^2}.
\end{equation}
In \eqref{SNR}, the received SNR depends on the beamforming vector $\mathbf{w}$ as well as the tilt angle $\theta$  and azimuth angle $\phi$ at the BS. In addition, it depends on the phase shifts $ {\Psi}= \left[ \psi_1, \psi_2 , \ldots, \psi_N\right] $ at the RIS. Since, we assume that there is only one user in the network, for any given set of $\theta$, $\phi$, and $ \mathbf{{\Psi}}$,   maximal ratio transmission (MRT) beamforming at the BS is optimal and hence, we set \cite{Wu} 
\begin{equation}
\mathbf{w}^*= \frac{\left(  \sqrt{\alpha^d } \ \mathbf{h}_d^H    
	+ \sqrt{\alpha^r} \ \mathbf{g}^H  \mathbf{\Omega} \  \mathbf{H}_r  \right)^H }{\left\| \sqrt{\alpha^d } \ \mathbf{h}_d^H    
	+ \sqrt{\alpha^r} \ \mathbf{g}^H  \mathbf{\Omega} \  \mathbf{H}_r \right\| }. 
\end{equation}

\section{Joint 3D Beamforming and Phase Optimization}
\label{sectionoptimization}
As we mentioned before, our main aim in this paper is to jointly optimize the 3D beamforming parameters (i.e., the tilt and azimuth angles at the BS) and the phase shifts $\mathbf{\Psi}$ at the RIS to have the best signal at the UE. We assume that  perfect channel state information (CSI) of $\mathbf{h}_d$, $\mathbf{H}$, and $\mathbf{g}$ are available at the BS. The objective of the optimization  is to maximize the received SNR at the UE as in \eqref{SNR}  subject to some constraints on the values of the tilt and azimuth angles as well as the phase shifts at the RIS elements. The resulting optimization problem can be formulated as 
\begin{equation} \label{optimization1}
\begin{aligned}
&\underset{\theta, \phi, \mathbf{\Psi}  }{\text{max}}&& \gamma(\theta,\phi, \Psi) = {   \left|  \left(\sqrt{\alpha^d } \ \mathbf{h}_d^H    
	+ \sqrt{\alpha^r} \ \mathbf{g}^H  \mathbf{\Omega} \  \mathbf{H}_r \ \right) \mathbf{w}^* \right|^2  }/{\sigma^2} \\
&\text{s.t.} &&   - \pi/2 \leq \theta \leq  \pi/2  \\
& \ &&  0 \leq \phi \leq  \pi \\
& \ &&   0 \leq {\psi_i} \leq 2\pi, \;  i=1, \ldots, N. \\
\end{aligned}
\end{equation}
This optimization problem is complex to solve it analytically but it can be solved numerically. In the next section, the results of this optimization problem are presented. 
\begin{figure}[t]
	\centering
	\begin{subfigure}[b]{0.48\textwidth}
		\centering
		\includegraphics[width=9cm,height=5.7cm]{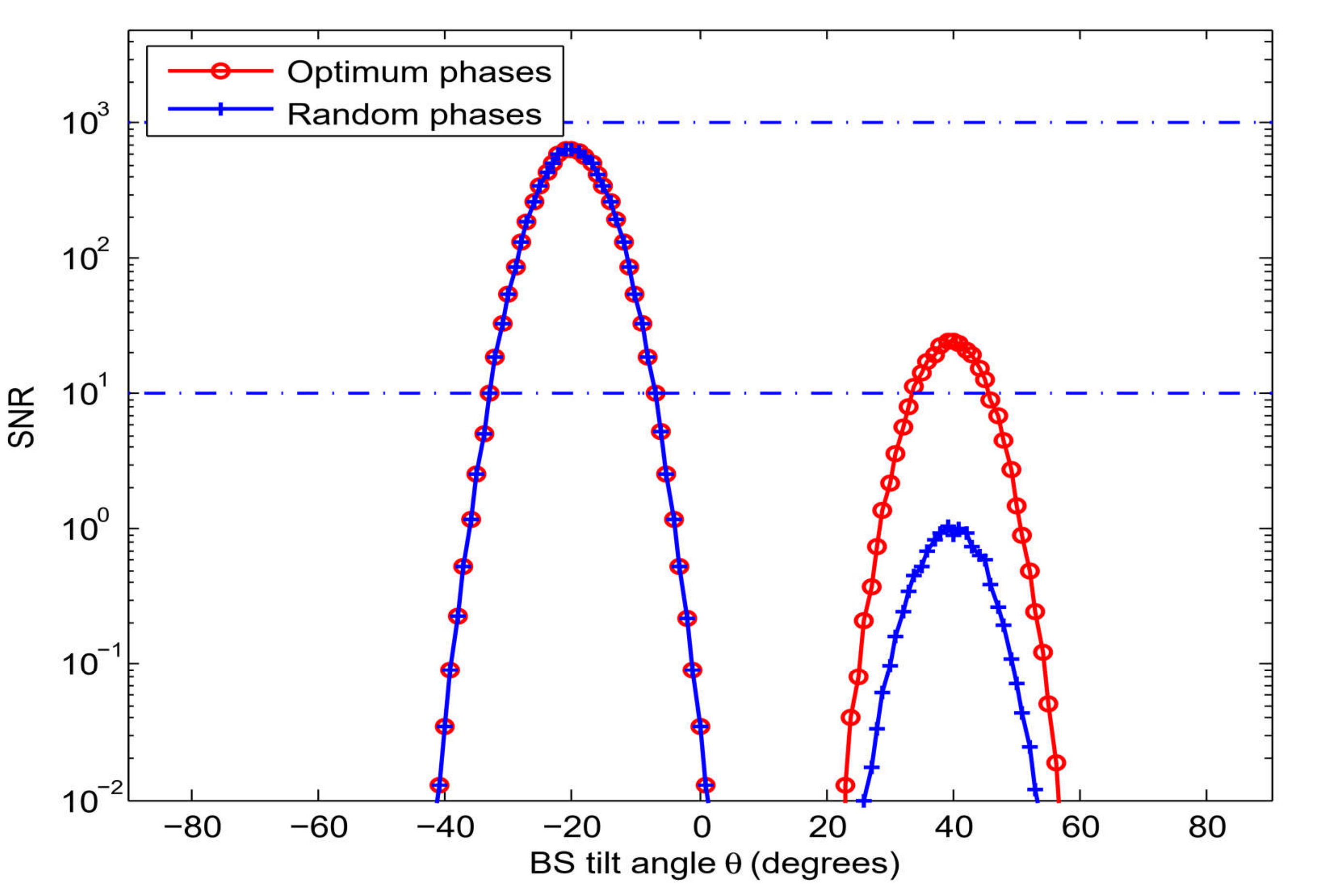}
		\caption{$\phi = 10^o$ \\ \ }
		\label{fig3phi10new}
	\end{subfigure} 
	\hfill
	\begin{subfigure}[b]{0.48\textwidth}
		\centering
		\includegraphics[width=9cm,height=5.7cm]{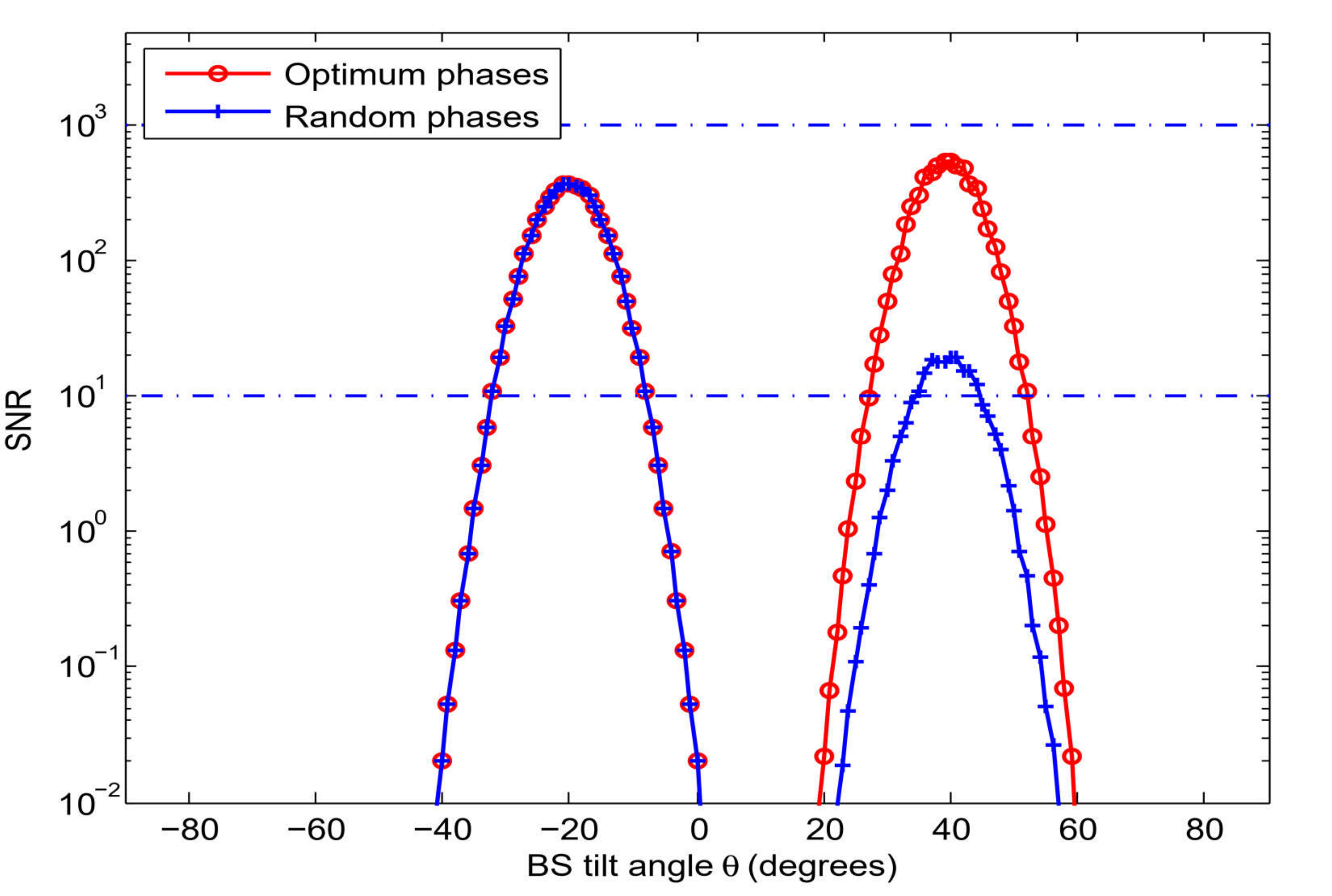}
		\caption{$\phi = 30^o$\\ \ }
		\label{fig3phi30new}
	\end{subfigure}   	
	\hfill
	\begin{subfigure}[b]{0.48\textwidth}
		\centering
		\includegraphics[width=9cm,height=5.7cm]{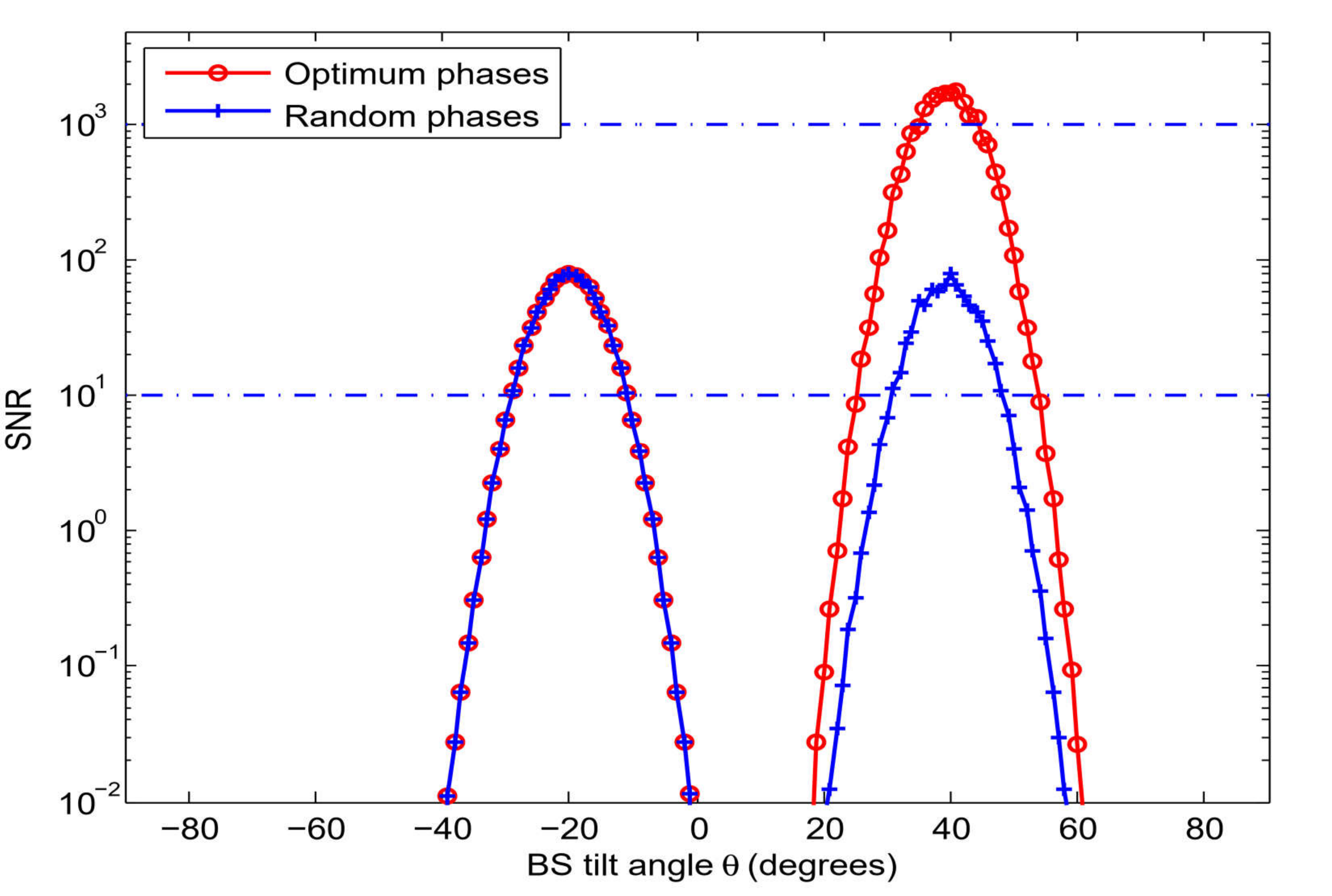}
		\caption{$\phi = 50^o$}
		\label{3phi50new}
	\end{subfigure}   
	\caption{\small{Received SNR at the UE versus the BS tilt angle for different BS azimuth angle $\phi$ (for optimum and  random phase shift at the RIS).}}
	\label{phi103050}
\end{figure}

\section{Simulation Results}
\label{sectionsimulation}


In this section, we show the results of our numerical simulations that are performed to evaluate the performance of the proposed scheme.  We assume that the BS is equipped with a planar array with $M=64$ antenna elements. The 3D parameters of the antenna array are set to $A_{max} = 0 $ dB, $\theta_{3dB}=15^o$ and $\phi_{3dB}=65^o$ \cite{Nadeem}.  The relative elevation and azimuth angle of the RIS with respect to the BS is $\theta_\mathrm{RIS}=40^o$ and $\phi_\mathrm{RIS}=50^o$. As it was shown in Fig. \ref{Fig3Dmodel}, these values depend on the relative position of the RIS and the BS, i.e. the relative height of the RIS to the BS and its distance to the BS. In addition, we assume that the UE is located in a position that its elevation and azimuth angles from the BS view are  $\theta_\mathrm{UE}=-20^o$ and $\phi_\mathrm{UE}=10^o$.

The optimum values of the parameters $\theta_\mathrm{BS}$, $\phi_\mathrm{BS}$ and  the RIS phase shifts $ \mathbf{\Psi}$ can be jointly obtained from \eqref{optimization1}. However, in our numerical simulations, to better show the effect of  tilt and azimuth angles on the performance, we first fix the tilt and azimuth angles and then for a given $\theta_\mathrm{BS}$ and  $\phi_\mathrm{BS}$, we find the optimum phase shift vector $\mathbf{\Psi}$ from \eqref{optimization1}. 

\begin{figure}[t]
	\centering
	\includegraphics[width=9cm,height=6.2cm]{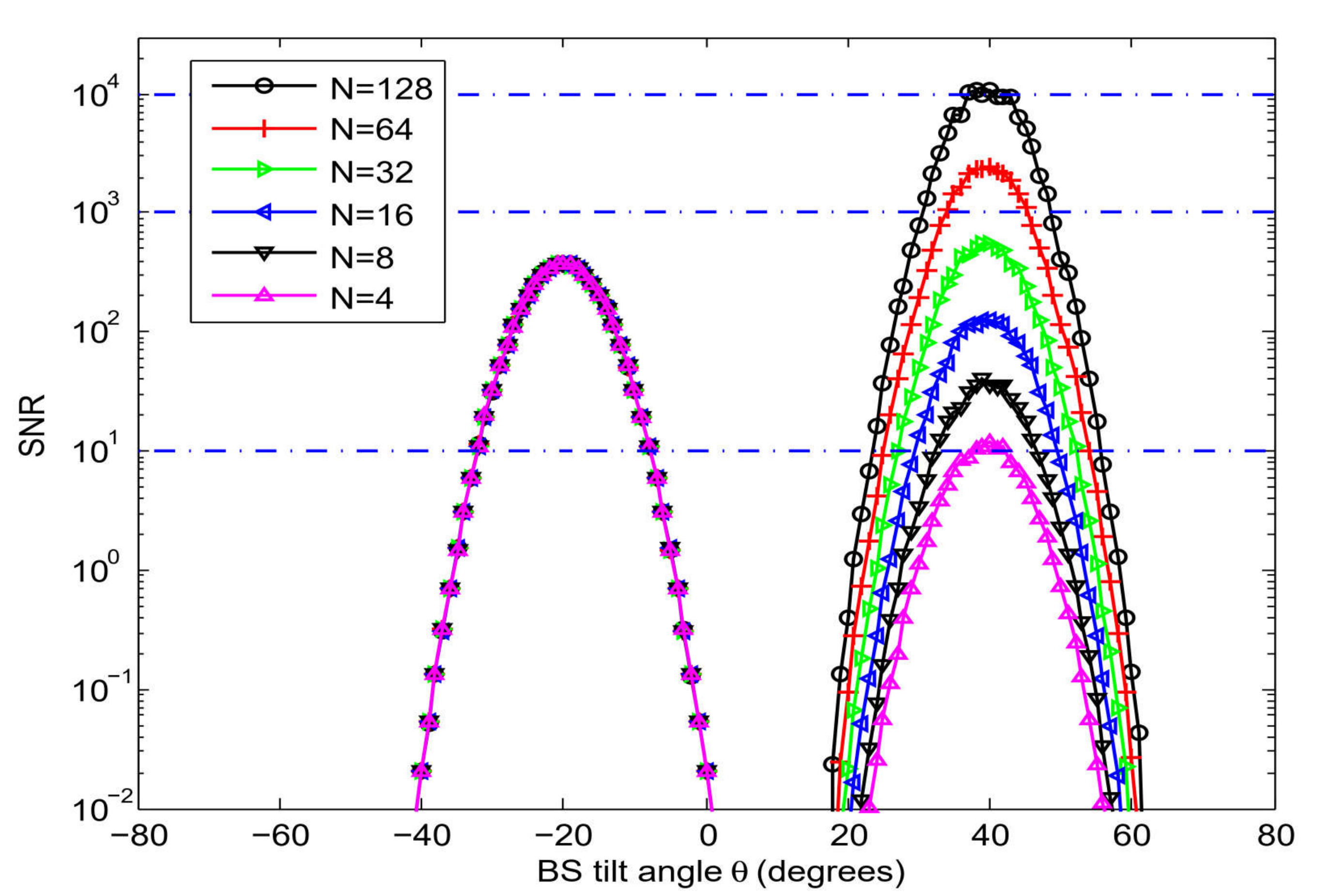}
	\caption{\small{Received SNR at the UE versus the BS tilt angle for  different number of reflecting elements at the RIS ($N$) when optimum phase shifts are used at the RIS (with $\phi=30^o$).}}
	\label{Noptimum}
\end{figure}
Fig. \ref{phi103050} depicts the received SNR at the UE versus the BS tilt angle for three different azimuth angles of $\phi = 10^o,30^o,50^o$, each is shown in one of the sub-figures. We see that the received SNR changes when the BS changes the orientation of the radiation beam and the maximum value of the SNR is obtained when the tilt angle is towards the user (i.e. $\theta = -20^o$)  or when it is directed towards the RIS   (i.e. $\theta = -40^o$). In each sub-figure, two graphs are presented which correspond to the optimum RIS phases (obtained from \eqref{optimization1}) and the random phase at the RIS. In all the cases, the optimum phase case shows substantial gains compared to the  random phase case. In addition,  when the BS radiating beam in the azimuth plane is towards the user (i.e. $\phi = 10^o$), the effect of the RIS phase optimization is less than the case in which the BS radiating beam in the azimuth plane is towards the RIS (i.e. $\phi = 50^o$). The best performance is when the radiated beam is towards the RIS in both azimuth and vertical planes and the phase shifts at the RIS are optimized through \eqref{optimization1}. This shows the effect of using RIS in the network. 

Fig. \ref{Noptimum} shows the effect of the number of RIS elements $N$ on the received SNR when the optimum phase shifts are used at the RIS. In this figure, the BS's azimuth angle is set to  $\phi = 30^o$. We see that when the radiating beam is towards the RIS, the performance improves with the number of reflecting elements. However, when the BS tilt is towards the user, the number of RIS elements does not have any considerable effect on the performance. In addition, it can be seen that the performance of the RIS-assisted network with a large number of reflecting elements ($N\geq32$) is much better than a  network that doesn't use RIS. 


\section{Conclusion}
\label{section conclusion}
In this paper, we presented a transmission scheme for wireless networks that  are empowered by an RIS. The proposed scheme adopts the 3D beamforming technique at the BS and tries to jointly optimize the tilt and azimuth angles of the BS and the phase shifts at the RIS. The optimum values of these parameters are obtained by numerical optimization and the effect of changing the BS radiation angles on the performance is studied. In addition, we studied the effect of the number of reflecting elements at the RIS taking into account the effect of angle of incidence of the received signal by the RIS on its reflecting properties. 
In summary, from the simulation results we conclude that when an RIS with a large number of reflecting elements and optimized phases is deployed at the network and the BS also uses 3D beamforming to optimize its radiation pattern direction, the performance of the network will be considerably better than the network that does not use RIS or an RIS-assisted network that does not use 3D beamforming. 

This paper assumed a basic model of one RIS and a single user system to show the advantages of using 3D beamforming in RIS-assisted networks. In our future work, we will extend the analyzed system to a multi-RIS and multi-user scenario. 

\bibliographystyle{ieeetr}
\bibliography{refs}

\end{document}